%% file: EC_GSM_IoT_Sync.tex
\documentclass[conference]{IEEEtran}

\newcommand{\gettitle}{EC-GSM-IoT Network Synchronization with Support for Large Frequency Offsets}

\usepackage[%
 utf8,
]{inputenc}
\usepackage{floatflt}
\usepackage{color}
\usepackage{cite}
\usepackage[normalem]{ulem}
\usepackage{multirow}
\usepackage{graphicx}
\usepackage{amsmath}
\usepackage{amsthm}
\usepackage{amsmath}
\usepackage{amssymb}
\usepackage{verbatim}
\usepackage{psfrag,array}
\usepackage{amsmath}
\usepackage{bm}
\usepackage{epstopdf}
\usepackage{algorithmic}
\usepackage{booktabs}
\usepackage{footnote}
\usepackage{url}
\usepackage{xcolor}
\usepackage{float}
\usepackage[nolist,printonlyused]{acronym}
\usepackage{threeparttable}
\usepackage{siunitx}
\usepackage[hidelinks,
            pdftitle={\gettitle}
            ]{hyperref}  

\newcommand{\p}[1]{\mathop{\mbox{\it p} } }

\newcommand{\be}{\begin{equation}}
\newcommand{\ee}{\end{equation}}
\newcommand{\ba}{\begin{array}}
\newcommand{\ea}{\end{array}}
\newcommand{\bea}{\begin{eqnarray}}
\newcommand{\eea}{\end{eqnarray}}
\newcommand{\bean}{\begin{eqnarray*}}
\newcommand{\eean}{\end{eqnarray*}}
\newcommand{\argmax}{\mathop{\arg\max}}

\newcommand{\ppm}{\text{ppm}}
\newcommand{\var}{\text{Var}}

\DeclareSIUnit[per-mode=symbol]\bit{b}
\DeclareSIUnit[per-mode=symbol]\byte{B}
\DeclareSIUnit[per-mode=symbol]\symbol{S}
\DeclareSIUnit[per-mode=symbol]\operation{OP}
\DeclareSIUnit\ppm{ppm}
\DeclareSIUnit\ppb{ppb}
\DeclareSIUnit\dbm{dBm}
\DeclareSIUnit{\nothing}{\relax}

\newcolumntype{L}[1]{>{\raggedright\let\newline\\\arraybackslash\hspace{0pt}}m{#1}}
\newcolumntype{C}[1]{>{\centering\let\newline\\\arraybackslash\hspace{0pt}}m{#1}}
\newcolumntype{R}[1]{>{\raggedleft\let\newline\\\arraybackslash\hspace{0pt}}m{#1}}

\definecolor{white}{rgb}{1,1,1}

\graphicspath{{./fig/}}

\newcommand{\figurewidth}{88mm}

\begin{document}

\bstctlcite{bibliography:BSTcontrol}

\title{\gettitle}

\author{
\IEEEauthorblockN{
Stefan~Lippuner\IEEEauthorrefmark{1},
Benjamin~Weber\IEEEauthorrefmark{1},
Mauro~Salomon\IEEEauthorrefmark{1},
Matthias~Korb\IEEEauthorrefmark{1},
Qiuting~Huang\IEEEauthorrefmark{1}
}
\IEEEauthorblockA{\IEEEauthorrefmark{1}Integrated Systems Laboratory, ETH Zurich, Zurich, Switzerland}
\{lstefan,weberbe,msalomon,mkorb,huang\}@iis.ee.ethz.ch
}

\IEEEoverridecommandlockouts
\IEEEpubid{\begin{minipage}[t]{\columnwidth}\copyright2018 IEEE. \newline Available online at https://ieeexplore.ieee.org/document/8377168/. Cite as: S. Lippuner, B. Weber, M. Salomon, M. Korb, and Q. Huang, “EC-GSM-IoT network synchronization with support for large frequency offsets,” in Wireless Communications and Networking Conference (WCNC), 2018. DOI: 10.1109/WCNC.2018.8377168. \end{minipage}\hspace{\columnsep}\makebox[\columnwidth]{ }}

\maketitle

\IEEEpubidadjcol

\input{content.tex}

\bibliographystyle{IEEEtran} \bibliography{IEEEabrv,./bibliography}

\end{document}

%% file: content.tex

\acresetall
\input{acronyms.tex}

\begin{abstract}
\ac{EDGE}-based \ac{EC-GSM-IoT} is a promising candidate for the billion-device \ac{cIoT}, providing similar coverage and battery life as \ac{NB-IoT}. The goal of \SI{20}{\decibel} coverage extension compared to \ac{EDGE} poses significant challenges for the initial network synchronization, which has to be performed well below the thermal noise floor, down to \iac{SNR} of \SI{-8.5}{\decibel}. We present a low-complexity synchronization algorithm supporting up to \SI{50}{\kilo\hertz} initial frequency offset, thus enabling the use of a low-cost \SI{\pm 25}{\ppm} oscillator. The proposed algorithm does not only fulfill the \ac{3GPP} requirements, but surpasses them by \SI{3}{\decibel}, enabling communication with \iac{SNR} of \SI{-11.5}{\decibel} or a maximum coupling loss of up to \SI{170.5}{\decibel}.
\end{abstract}

\acresetall

\input{acronyms.tex}
\section{Introduction}
Cellular networks have enabled five billion people to connect to the Internet. This number is expected to rise further, with \ac{4G} and \ac{5G} standards meeting the high throughput and low latency requirements of mobile phones. In addition to this, it is estimated that billions of devices will also be directly connected to the Internet. The \ac{3GPP} has developed three new standards to address the needs of this \acf{IoT}: \ac{EC-GSM-IoT}, \ac{LTE} Cat-NB (\acs{NB-IoT}), and \ac{LTE} Cat-M (\acs{eMTC}) all provide better coverage and reduce both cost and power consumption for low-rate devices \cite{3gpp:rel13}. 

The vast majority of \ac{cIoT} devices currently use \ac{GPRS}/\ac{EDGE} to connect to the Internet \cite{nokiaIoTConn}. For these, \ac{EC-GSM-IoT} provides an easy path to improved energy efficiency and a \SI{20}{\decibel} coverage improvement. \ac{EC-GSM-IoT} specifies \SI{164}{\decibel} of \ac{MCL} and an expected battery life of more than 10 years, matching the other \ac{cIoT} standards. Additionally, \ac{EDGE} support provides instantaneous global coverage and allows the throughput to scale up to \SI[per-mode=symbol,per-symbol = p]{355}{\kilo\bit\per\second}, comparable to half-duplex \ac{LTE} Cat-M1. Further, \ac{EC-GSM-IoT} is expected to have the lowest module cost of the \ac{3GPP} standards \cite{nokiaIoTConn}.

The problem of synchronizing to a legacy \ac{GSM} carrier has been extensively studied and several low-complexity solutions exist \cite{kroll2012low, jha2000acquisition}. For \ac{EC-GSM-IoT}, however, the sensitivity requirement is \SI{20}{\decibel} more stringent, and the signal level is now well below the thermal noise floor. The legacy algorithms are unable to cope with this, and new approaches are required to synchronize at \acp{SNR} as low as \SI{-8.5}{\decibel}. 

In this paper, we present an algorithm, which is able to perform the \ac{EC-GSM-IoT} network synchronization down to \iac{SNR} of \SI{-11.5}{\decibel} in a low-band \ac{ST} channel. The support for frequency offsets of up to \SI{50}{\kilo\hertz} allows the use of a \SI{\pm 25}{\ppm} crystal oscillator and, therefore, enables a low overall module cost. Compared to \cite{weber2017sawless}, the \ac{MF} boundary can be detected at \SI{-14.4}{\decibel} instead of \SI{-11}{\decibel} \ac{SNR} in \iac{ST} channel, and the average synchronization time at \SI{-8.5}{\decibel} \ac{SNR} has been reduced by \SI{56}{\percent} to \SI{1.3}{\second}. The proposed \ac{FOE} can accurately estimate the frequency offset at \SI{-8.5}{\deci\bel} \ac{SNR}.

The remainder of this paper is structured as follows: Section~\ref{sec:ec-gsm network sync} introduces the \ac{EC-GSM-IoT} \ac{MF}, the proposed synchronization procedure, the system model, and the synchronization requirements. The proposed algorithms are discussed in detail in Section~\ref{sec:algorithm}. Finally, the performance for synchronizing to a known carrier and the cell search are evaluated in Sections \ref{sec:performance} and \ref{sec:cell search}.


\section{\acs{EC-GSM-IoT} Network Synchronization}
\label{sec:ec-gsm network sync}

\subsection{\acl{MF} Structure}

\begin{figure*}[ht]
\begin{center}
\includegraphics[width=\textwidth]{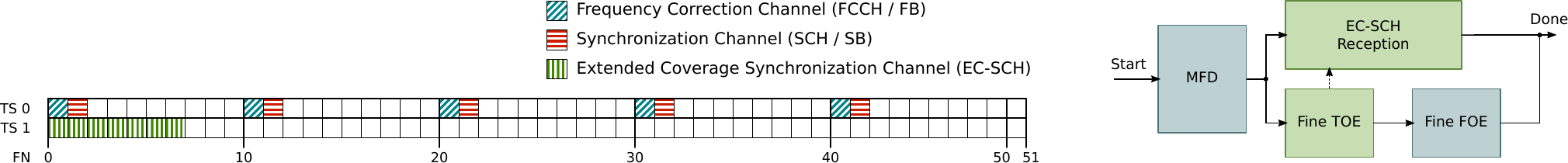}
\caption[EC-GSM-IoT Multiframe]{Left: \acs*{EC-GSM-IoT} \acs*{MF}. In addition to the legacy \acsp*{FB} and \acsp*{SCH}, seven blind repetitions of the \acs*{EC-SCH} are present on \acs*{TS} 1 of the frames \numrange{0}{6}. \acp*{TS} \numrange{2}{7} are not shown, since they are not used for the synchronization. Right: Synchronization flow for the proposed algorithm.}
\label{fig:EC-GSM-IoT Multiframe}
\end{center}
\end{figure*}

Fig.~\ref{fig:EC-GSM-IoT Multiframe} shows the elements in the \ac{EC-GSM-IoT} \ac{MF}, which enable the network synchronization. This \SI{235}{\milli\second} long structure is repeated on the \ac{BCCH} carrier. It consists of \num{51} frames with eight \acp{TS} each. Five \acp{FB} are transmitted on \ac{TS} 0 of frames 0, 10, 20, 30, and 40. The legacy \ac{SCH} follows exactly one frame later and carries the current frame number and the \ac{BS} identity code. Devices may not be able to decode the \ac{SCH} below the noise floor, and can use the \ac{EC-SCH} with 28 blind repetitions over four \acp{MF} instead. The presence of the \ac{EC-SCH} indicates that the networks supports \ac{EC-GSM-IoT}.

\subsection{Synchronization Procedure}


The proposed algorithm performs the synchronization in four sequential steps as shown in Fig.~\ref{fig:EC-GSM-IoT Multiframe}. First, the \ac{MFD} finds a coarse estimate for the start time of the \ac{MF} and the frequency offset, using the \acp{FB}. After the frequency offset has been corrected for, the fine time offset is estimated using the \ac{EC-SCH} training sequences. At the same time, the decoding of the \ac{EC-SCH} is started. The fine \ac{FOE} can be started, as soon as the fine time offset is known. The device is ready to receive the \ac{EC-BCCH}, once the \ac{EC-SCH} has been successfully decoded, and may transmit, once the fine \ac{FOE} is completed.

\subsection{System Model}

We use the same noise figure as the \ac{3GPP} reference document: $\text{NF}=\SI{5}{\decibel}$ \cite{3gpp:45820}. The noise bandwidth is $\text{BW} = \SI{200}{\kilo\hertz}$ and the sampling frequency is $f_s = \SI{270.8}{\kilo\hertz}$. The receiver is powered on at a random time, and the initial frequency offset is uniformly distributed between $\pm f_{o, \max} = \pm R f_c$. $R$ is the error relative to the carrier frequency $f_c$. The simulations for the low bands were performed at $f_c = \SI{900}{\mega\hertz}$ and at $f_c = \SI{2}{\giga\hertz}$ for the high bands.
\acused{NF}

Since the oscillator is a significant factor in the system cost, it is desirable to use a low-cost crystal oscillator for \iac{cIoT} application. These typically have an initial accuracy of $R = \SI{25}{\ppm}$. In the high bands, this corresponds to a maximum frequency offset of \SI{\pm 50}{\kilo\hertz}. This offset can be corrected by tuning the oscillator, or compensated for in the digital baseband.

\subsection{Synchronization Requirements}

A device is synchronized to the network, if it has successfully decoded the \ac{EC-SCH}, achieved timing synchronization, and corrected for the offset of the local oscillator. The requirements from the \ac{EC-GSM-IoT} standard are as follows \cite{3gpp:rel13}:

\begin{description}
\item[R1] The device must be able to synchronize to the \ac{BCCH} carrier within 2 seconds after power-on, which corresponds to successfully decoding the \ac{EC-SCH}.
\item[R2a] Before the first transmission, the timing offset with respect to the \ac{BS} must be below $\pm 1$ symbol periods.
\item[R2b] Before the first transmission, the relative frequency offset with respect to the \ac{BS} must be below \SI{\pm 0.1}{\ppm}.
\end{description}

Requirement~R1 has to be met at the \ac{ISL} for reference performance of the \ac{EC-BCCH}, while Requirements R2a and R2b have to be met at the \ac{ISL} for reference performance of the \ac{EC-SCH}. These requirements differ for the low and high frequency bands, and are specified for three radio channels conditions: \ac{ST}, \ac{TU}1.2, and \ac{TU}50. The latter are fading with a mobile speed of \SI[per-mode=symbol]{1.2}{\kilo\meter\per\hour} and \SI[per-mode=symbol]{50}{\kilo\meter\per\hour}. The \ac{EC-SCH} and \ac{EC-BCCH} \ac{ISL} for the \ac{ST} channels correspond to \iac{SNR} of \SI{-8.5}{\decibel} and \SI{-7.5}{\decibel}. The \acp{SNR} for the low band \ac{TU}1.2 channels are \SI{-5.5}{\decibel} and \SI{-3.5}{\decibel} respectively. Since no maximum miss rate is specified for Requirement~R1, we will use the same as specified for the \ac{EC-SCH} \ac{BLER}: \SI{10}{\percent} misses.

The \SI{2}{\second} from Requirement~R1 equal approximately eight \acp{MF}. In the worst-case, it takes seven \acp{MF} to receive a complete set of 28 blind \ac{EC-SCH} repetitions. Therefore, the \ac{MFD} should usually only require a single \ac{MF}. If the \ac{MFD} takes more than four \acp{MF}, it is no longer possible to collect all 28 blind repetitions of the \ac{EC-SCH} within \SI{2}{\second}.


\section{Network Synchronization Algorithm}
\label{sec:algorithm}

\subsection{Fine Frequency Offset Estimation}
The fine \ac{FOE} is performed as the last step of the synchronization. This allows us to discuss it as a stand-alone problem, assuming perfect timing synchronization. The \ac{FOE} estimates the offset of the local oscillator compared to the \ac{BS} using the regularly transmitted \acp{FB}. Each \ac{FB} consists of a pure sine at $f_s /4 = \SI{67.7}{\kilo\hertz}$ for \num{148} symbol durations. The problem of estimating the frequency of a sine in \ac{AWGN} is a well-studied problem, and the \ac{CRLB} has been derived in \cite{rife1974single}. We have adapted it to the \ac{GSM} noise bandwidth:

\begin{equation} \label{eq:ml_crlb_b}
\var(\hat{\omega}) \geq \frac{6 \cdot f_s{}^3}{\text{SNR} \cdot \text{BW} \cdot N (N^2 -1)}. \\
\end{equation}

The \ac{ML} estimator for this problem selects the frequency with the maximum power spectral density $|A(f)|^2$:

\begin{equation} \label{eq:ml}
\hat{f} = \argmax_f |A(f)|^2.
\end{equation}

It approaches the \ac{CRLB} for high \ac{SNR} and can be approximated using \iac{DFT} \cite{rife1974single}.

\begin{figure}[!ht]
\begin{center}
\includegraphics[width=\figurewidth]{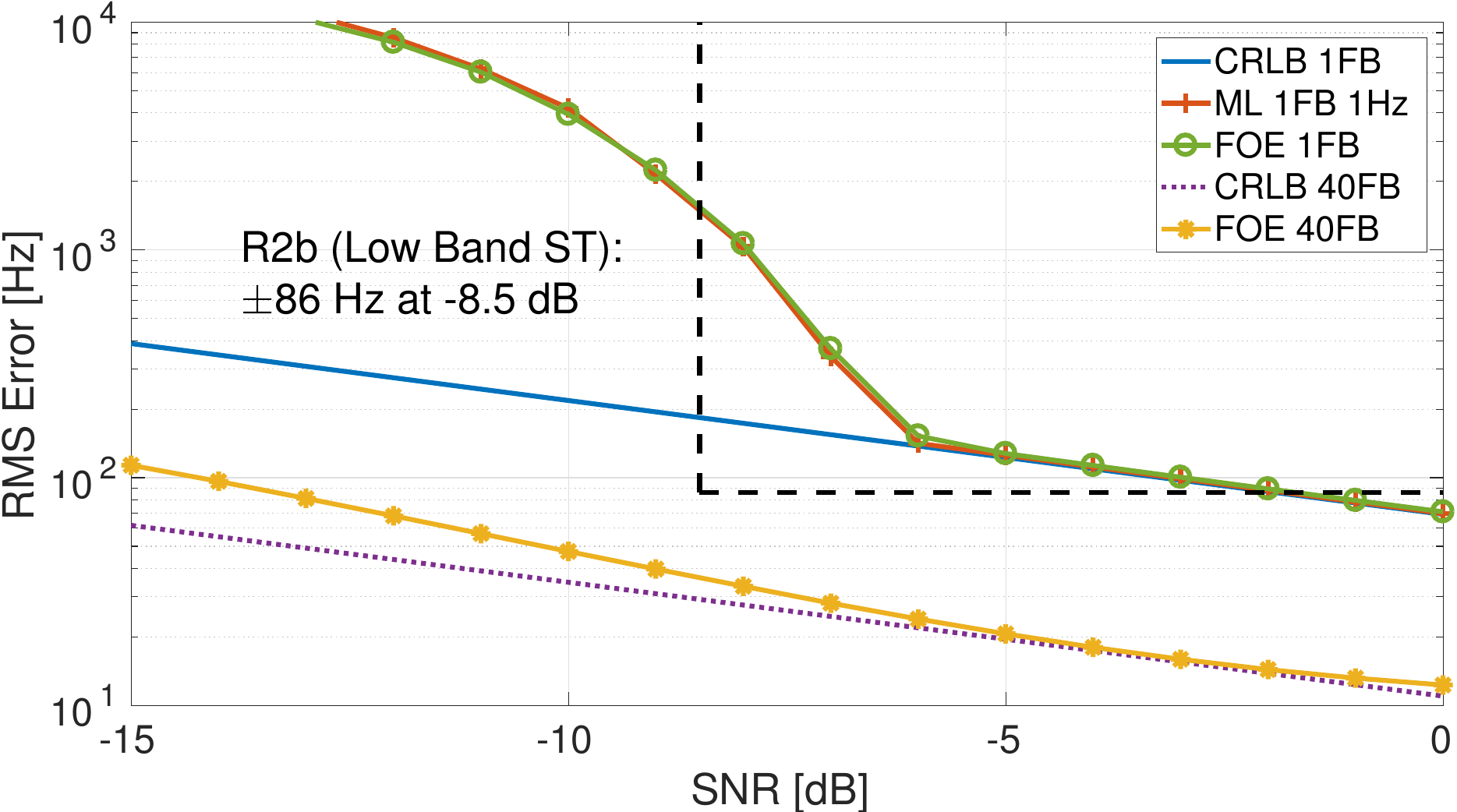}
\caption[FOE combination methods: RMS error]{\acs*{RMS} frequency offset improvement by considering 40 \acsp*{FB} in a static channel. The \acs*{CRLB} is shown twice, once for a single \acs*{FB} and once with a 40 times larger receive power. The ML estimator is approximated using a \SI{1}{\hertz} resolution \acs*{DFT}.}
\label{fig:FOE Combination Methods}
\end{center}
\end{figure}

To fulfill Requirement~R2b at the minimum carrier frequency of \SI{869}{\mega\hertz}, the frequency has to be estimated with a maximum error of \SI{86}{\hertz}. At \iac{SNR} of \SI{-8.5}{\decibel}, the \ac{CRLB} for the \ac{RMS} error is \SI{182}{\hertz}, as is shown in Fig.~\ref{fig:FOE Combination Methods}. Clearly, this is not sufficient to achieve the required performance. The \ac{ML} estimator also suffers from a threshold effect, which renders it useless below \iac{SNR} of \SI{-6}{\decibel}.

The proposed algorithm combines the information from several \acp{FB} in order to improve the accuracy of the estimate. The phase of the \acp{FB} does not provide any information, since \ac{GSM} has no guaranteed phase coherency between frames. $N$ \acp{FB} are used by non-coherently accumulating the power of the individual \acp{FB} in the frequency domain:

\begin{equation} \label{eq:foe_comb_nc}
\hat{f} = \argmax_f A'(f), \qquad A'(f) = \sum\limits_{i=0}^{N-1} |A_{i}(f)|^2.
\end{equation}

Our simulations show that this non-coherent combination achieves almost the same performance as a coherent combination with perfect knowledge of the phase.

The \ac{ML} estimator can be approximated using a two-step procedure to keep the implementation complexity low. In the first step a 256-point \ac{FFT} and the power for the frequency bins of interest are calculated. At this point, the discrete spectrum can be accumulated over multiple \acp{FB}. The maximum of these \ac{FFT} bins is then taken as the coarse location of the spectral peak. In a second step, a three-point interpolation is applied to produce a higher resolution estimate of the sine frequency. The algorithm from \cite{yang2011noniterative} calculates a correction term $f_a$ using only the power in the three \ac{DFT} bins, $Y_{-1}, Y_0,$ and $Y_1$, closest to the spectral peak:

\begin{equation} \label{eq:foe_rctsl}
f_{\alpha} = 2\pi \cdot \frac{|Y_1|^2 - |Y_{-1}|^2}{u(|Y_1|^2 + |Y_{-1}|^2) + v|Y_0|^2}.
\end{equation}

The original algorithm in \cite{yang2011noniterative} requires \iac{DFT} size of $2N$. We have modified the derivation of the constants $u$ and $v$ in order to allow for an arbitrary \ac{DFT} size, such that \iac{FFT} can be used:

\begin{align} \label{eq:foe_rctsl_l}
u & = \frac{sin^2 \left( \frac{\pi N}{L} \right) \left[ \sin^2 \left( \frac{\pi N}{L} \right) \sin(\frac{2\pi}{L}) 
- N \sin \left( \frac{2 \pi N}{L}\right) \sin^2 \left( \frac{\pi}{L} \right) \right]}
{\sin^2 \left( \frac{\pi}{L} \right) \left[ N^4 \sin^4 \left( \frac{\pi}{L} \right) + 
2 \sin^4 \left( \frac{\pi N}{L} \right) \right] }, \\
v &= u \cdot \frac{N^2 \sin^2 \left( \frac{\pi}{L} \right)}{\sin^2 \left( \frac{\pi N}{L} \right)},
\end{align}
where $N$ is the number of samples and $L$ is the \ac{DFT} size.

Fig.~\ref{fig:FOE Combination Methods} shows the simulated performance of the proposed \ac{FOE} algorithm in \iac{ST} channel, compared to the \ac{ML} estimator and the \ac{CRLB}. At the target \ac{SNR} of \SI{-8.5}{\decibel}, the \ac{RMS} frequency estimation error is \SI{36}{\hertz}, clearly meeting Requirement~R2b. Once the frequency of the sine has been estimated, it is also possible to estimate the amplitude of the sine from the same \ac{DFT} bins. This can be used to detect the presence of \iac{FB} and the \ac{MF} boundary.

\subsection{Multi-Frame Boundary Detection}

The \ac{MFD} is the first step in the synchronization process. Its main purpose is to estimate the start time of the \ac{MF}, and therefore the position of the \ac{EC-SCH}. One option is the use of \iac{ML} estimator, like the one proposed in \cite{kroll2017maximum} for \ac{NB-IoT}. However, the complexity of such an estimator scales with the maximum initial frequency offset and the number of correlated signals. For \ac{EC-GSM-IoT} the FCCH and the training sequences of the \ac{SCH} and \ac{EC-SCH} are known a-priori and can be used for the search. Based on \cite{kroll2017maximum}, we estimate the complexity for \ac{EC-GSM-IoT}: Three different signals need to be cross-correlated with, and the maximum frequency offset is three times larger than in \cite{kroll2017maximum}, resulting in a nine times larger real arithmetic complexity of \SI[per-mode=symbol]{2.5}{\giga\operation\per\second}. Storing the 9-bit correlation for \num{96} frequency and \num{63750} time offset candidates requires \SI{55}{\mega\bit} of memory. This is clearly not suited for a low-power and low-cost implementation and we therefore propose a reduced complexity \ac{MFD} algorithm.

The proposed \ac{MFD} algorithm processes the received data in sliding windows of 200 symbols, which are offset by 50 symbols. The \ac{FOE} procedure is re-used to look for the maximum spectral component in each window. The amplitude at this frequency, $C_{\text{FB}}[n]$, and the frequencies of the two highest components are stored. If the window contains \iac{FB}, there is a high chance that one of these two frequencies corresponds to the sine. Since \num{1275} windows, corresponding to one \ac{MF}, are considered, \num{3825} elements have to be stored in memory. Once an entire \ac{MF} worth of samples has been received, the algorithm accumulates the five \ac{FB} correlations for \iac{MF} start in each window:

\begin{equation} \label{eq:mf_corr}
C_{\text{MF}}[n] = \sum\limits_{i=0}^{4} C_{\text{FB}}[n + 250i \mod 1275].
\end{equation}

\begin{figure}[!ht]
\begin{center}
\includegraphics[width=\figurewidth]{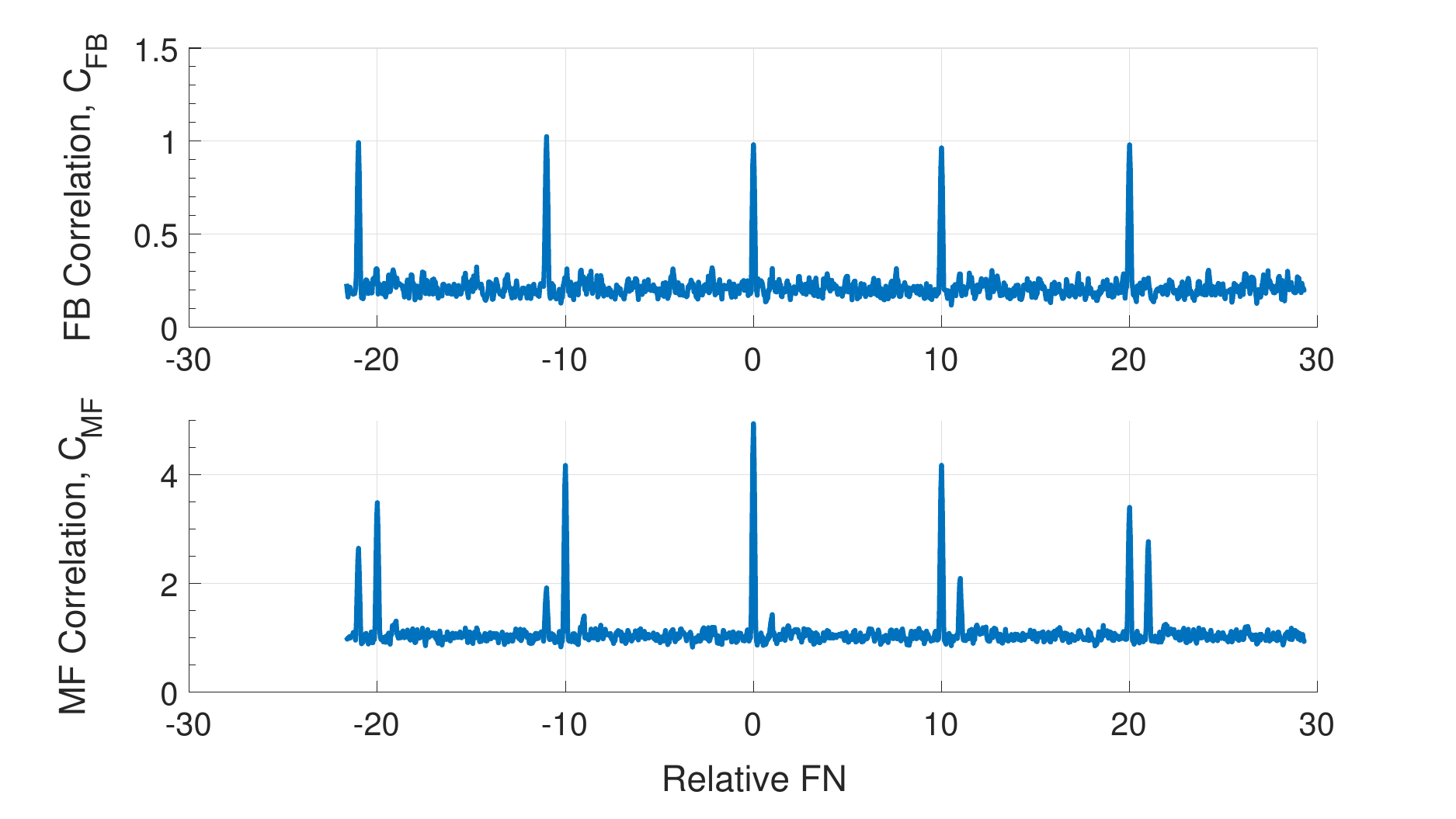}
\caption[MFD: Correlation with the FBs]{MFD: Correlation with the \acsp*{FB} and resulting \acs*{MF} correlation in \iac*{ST} channel without noise. The \acs*{MF} starts at $\text{FN} = 0$.}
\label{fig:MFD: Correlation with the FBs}
\end{center}
\end{figure}

Fig.~\ref{fig:MFD: Correlation with the FBs} shows the two metrics in the noiseless case. Note that the normal bursts still result in a non-zero \ac{FB} correlation.

The stored frequencies for the maximum of $C_{MF}$ are then compared. If there is a set of frequencies with a spread below an empirical threshold of \SI{1}{\kilo\hertz}, the search is considered successful. Otherwise, additional \acp{MF} are searched and $C_{\text{MF}}$ is accumulated and the frequencies are updated. On a hit, the frequencies are also used as a coarse estimate of the frequency offset. At \iac{SNR} of \SI{-8.5}{\decibel} in \iac{ST} channel, the residual \ac{RMS} frequency offset is \SI{150}{\hertz}, which is sufficient to continue with decoding the \ac{EC-SCH}.

One of the main challenges for the \ac{MFD} is the structure of the \ac{GSM} \ac{MF}. The quasi-regular spacing of the \acp{FB} implies that the only difference between two \ac{MF} starts is the location of a single \ac{FB}. This results in a number of false side-peaks in the \ac{MF} correlation, as can be seen in Fig.~\ref{fig:MFD: Correlation with the FBs}. This is especially problematic for fast fading channels, like the TU50 test case. We propose the use of the \ac{EC-SCH} training sequences to find the actual start of the \ac{MF}, since they follow a different repetition scheme. To this end, the \ac{MFD} does not only return the peak of $C_{MF}$, but also the two positions 10 frames earlier and later as possible \ac{MF} start candidates. The correct candidate is later found using the \ac{EC-SCH} training sequences.

\begin{figure}[!ht]
\begin{center}
\includegraphics[width=\figurewidth]{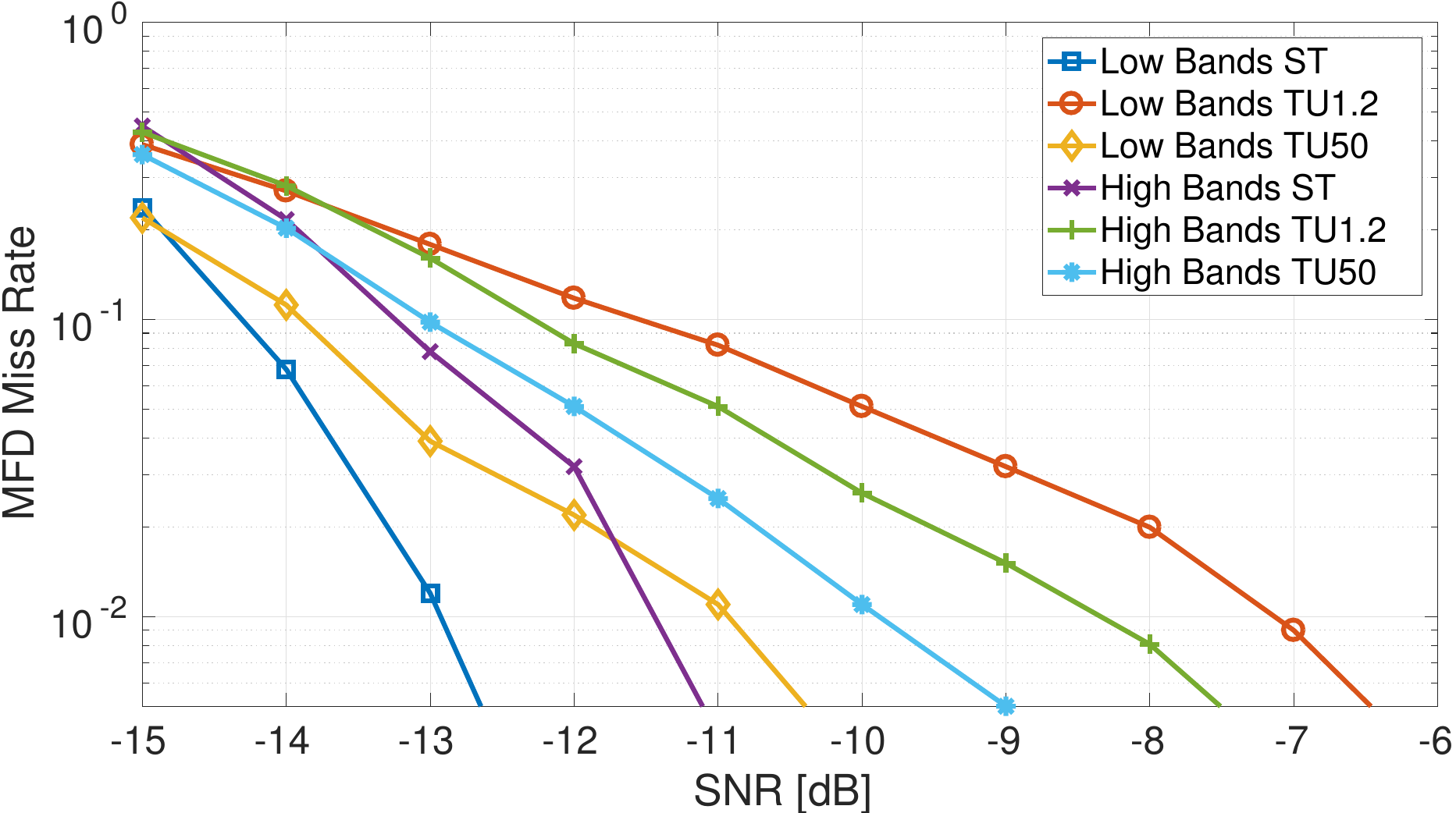}
\caption[MFD: Spectrum method miss rate for the EC-GSM-IoT channels]{Spectrum \acs*{MFD} method miss rate for the \acs*{EC-GSM-IoT} channels. If none of the three candidates are within \num{\pm 80} symbols of the \acs*{MF} start, the attempt is considered \iac*{MFD} miss.}
\label{fig:MFD: Spectrum method miss rate for the GSM channels}
\end{center}
\end{figure}

Fig.~\ref{fig:MFD: Spectrum method miss rate for the GSM channels} shows the performance of the Spectrum \ac{MFD} for the \ac{EC-GSM-IoT} channels in the low and high bands. In order to allow the \ac{EC-SCH} decoding sufficient time, the \ac{MFD} is terminated after searching 4 \acp{MF} and the most likely candidate is used.


The \ac{MFD} is the computationally most demanding part of the synchronization process. \num{5400} windows need to be processed every second, each requiring a 256-point \ac{FFT}. Including the calculation of the power and three-point interpolation, this results in an overall real arithmetic complexity of \SI[per-mode=symbol]{40}{\mega\operation\per\second}. This is suitable for a low-complexity hardware implementation and should result in a negligible power consumption, compared to the \ac{RF} receiver.

\subsection{Fine Time Offset Estimation and EC-SCH Decoding}

Once the \ac{MFD} is completed, the time offset is known to within \num{\pm 80} symbols. In a next step, the accuracy has to be improved to \num{\pm 1} symbol in order to decode the \ac{EC-SCH} and fulfill Requirement~R2a. This is done by cross-correlating the received \ac{EC-SCH} training sequences with the known symbols. The time offsets in the range of the residual time offset of \num{\pm 80} symbols are tried and the best match is selected. In a static channel at \iac{SNR} of \SI{-8.5}{\decibel}, the correct time offset is found in approximately \SI{60}{\percent} of all attempts. Blind repetitions of the training sequences are thus received and the correlation outputs $K_p[\eta]$ are combined non-coherently:

\begin{equation}
\hat{\eta} = \argmax_{\eta} K'[\eta], \quad K'[\eta] = \sum\limits_p |K_p[\eta]|^2.
\end{equation}

Seven blind repetitions of the \ac{EC-SCH} are sufficient to determine the correct time offset at \iac{SNR} of \SI{-8.5}{\decibel} in \iac{ST} channel. The correct \ac{MF} start candidate is selected by choosing the one with the largest $K'[\hat{\eta}]$ after \num{28} blind repetitions have been received.

The last steps in the synchronization procedure are the fine frequency offset estimation and the decoding of the \ac{EC-SCH}, which can be performed simultaneously. The \ac{EC-SCH} decoding is started together with the fine time offset estimation and uses the most recent result of the latter. Seven repetitions of the \ac{EC-SCH} are transmitted at the start of every \ac{MF} and a total of 28 repetitions contain the same data. Repetitions in different \acp{MF} however, have different cyclic shifts, which depend on the frame number. 

The actual decoding of the \ac{EC-SCH} is implemented, as described in \cite{weber2017sawless}, where the \acp{LLR} for the \ac{EC-SCH} repetitions are chase-combined for the four cyclic shift candidates. \ac{IQ}-combination can significantly improve the channel estimation, especially when short training sequences are used. However, the \ac{EC-SCH} uses longer training sequences and the phase between repetitions has to be estimated. Our simulations show that \ac{IQ}-combination does not improve the performance of the decoder due to the phase estimation error at \iac{SNR} of \SI{-8.5}{\decibel}. 

One option to detect the cyclic shift is comparing the data received in two successive \acp{MF} using a cross-correlation with lag \num{1}. Unfortunately, the correlation output is dominated by noise in the target \ac{SNR} region, and this method fails. Alternatively, blind decoding attempts can be performed for all cyclic shifts, relying on the \ac{CRC} checksum to filter out the invalid results. But this method also has a flaw, because the decoding attempts on invalid data can result in false positives from the checksum. The problem is further complicated by the relatively short \ac{CRC} field length of 10 bits. If decoding is attempted after every received repetition, \ac{CRC} false positives occur up to \SI{19}{\percent} of the time. The proposed algorithm alleviates this problem by reducing the number of decoding attempts. In the low \ac{SNR} region, a decoding attempt is only performed after $7N, N \geq 4$ repetitions have been received. This reduces the maximum false positive rate to \SI{2.7}{\percent}. It is below \SI{1}{\percent} above \SI{-8.5}{\decibel} \ac{SNR}, which is acceptable and has to be handled after the synchronization.


\section{Performance Evaluation}
\label{sec:performance}

\begin{figure}[!ht]
\begin{center}
\includegraphics[width=\figurewidth]{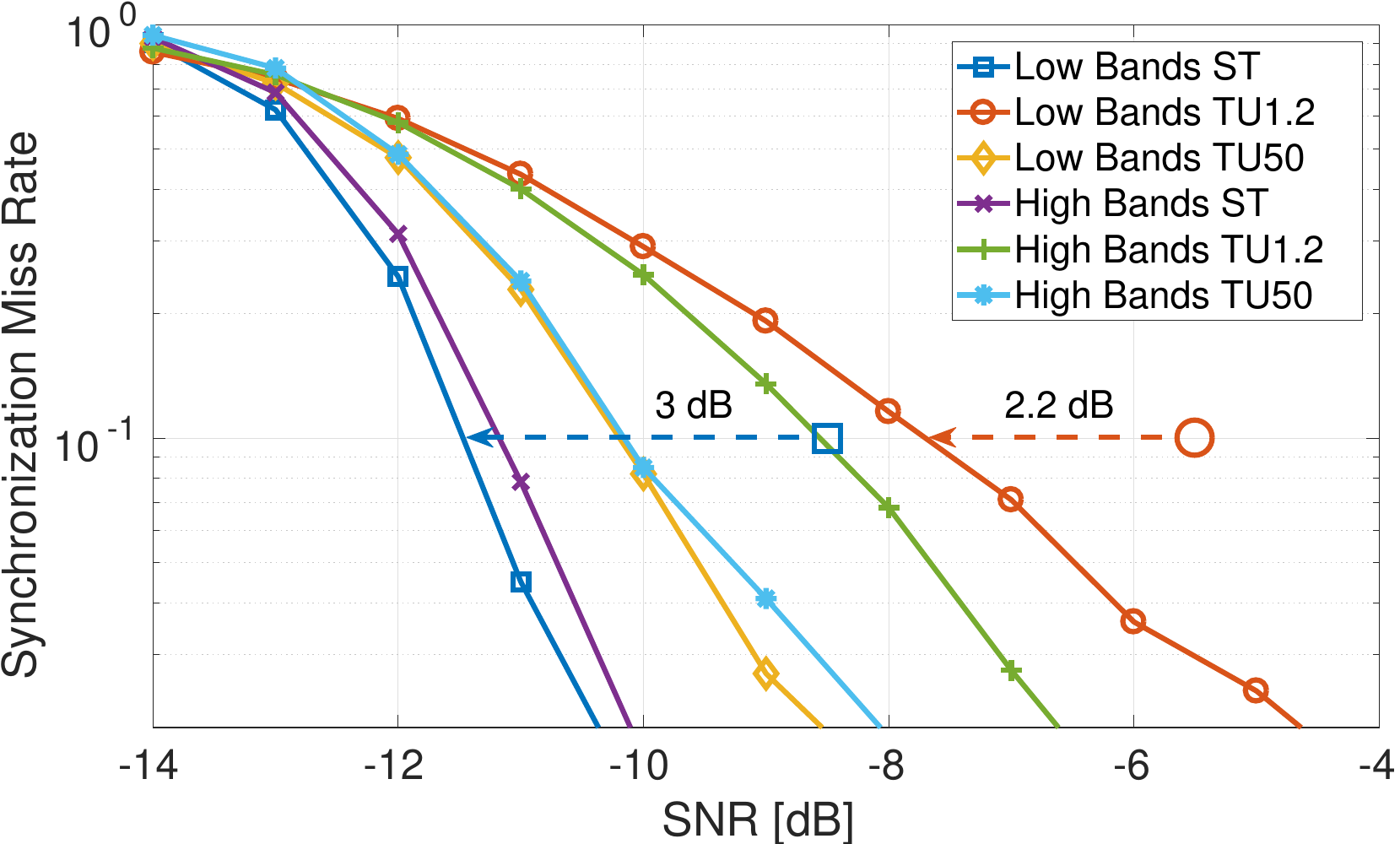}
\caption[Synchronization miss rate]{Synchronization miss rate for the different \ac*{EC-GSM-IoT} channels. A synchronization miss is either a timeout, or \iac*{CRC} hit with incorrect data. The initial oscillator error is uniformly distributed between \SI{\pm25}{\ppm}.}
\label{fig:Synchronization miss rate}
\end{center}
\end{figure}

\begin{figure}[!ht]
\begin{center}
\includegraphics[width=\figurewidth]{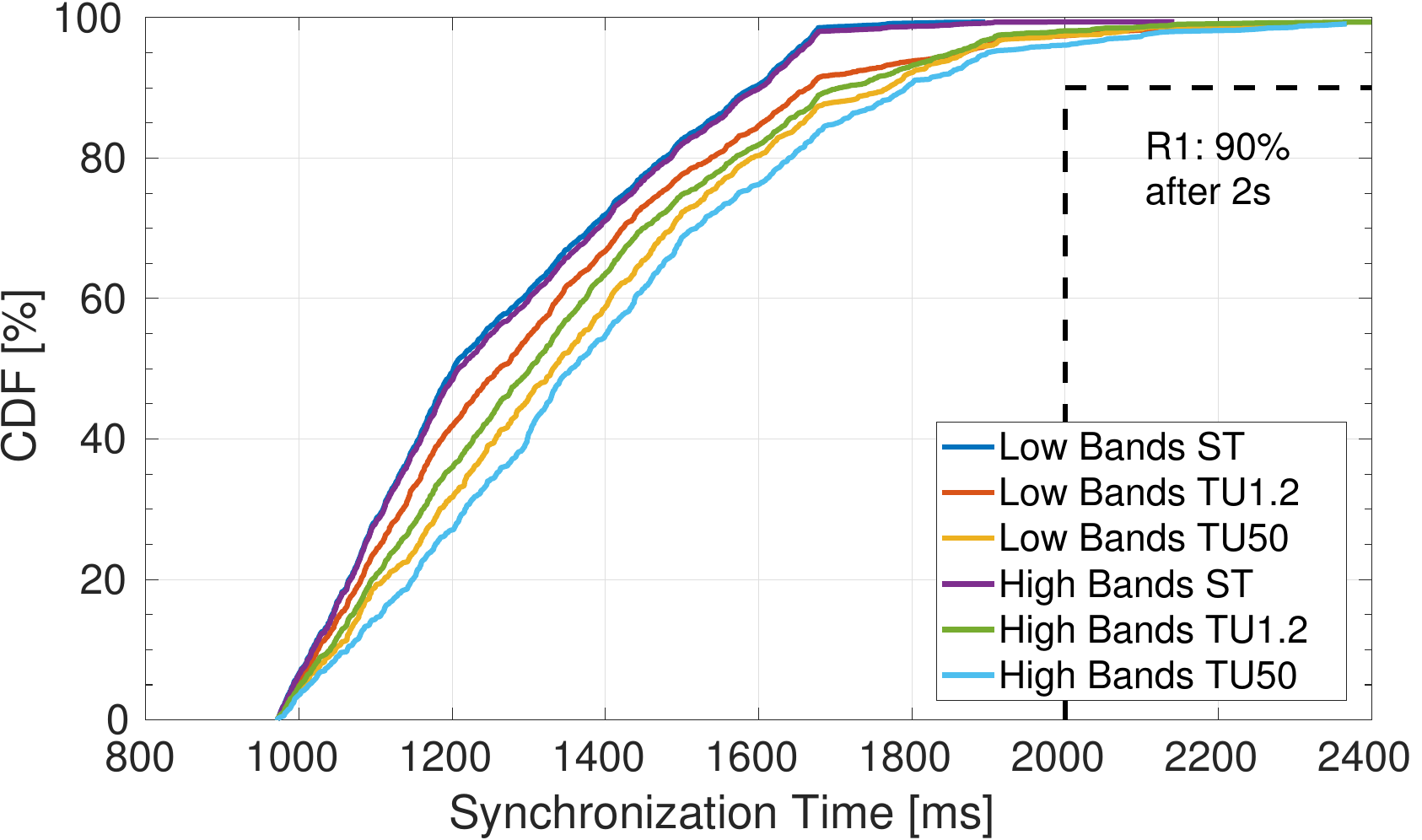}
\caption[Synchronization time CDF]{Synchronization time \ac*{CDF} at the \ac*{ISL} for \ac*{EC-BCCH} reference performance, assuming \iac*{NF} of \SI{5}{\decibel}. Unsuccessful synchronizations are counted towards to total.}
\label{fig:Synchronization time CDF}
\end{center}
\end{figure}

In order to evaluate the performance of the synchronization, we have performed simulations, where all the steps are performed in sequence. Fig.~\ref{fig:Synchronization miss rate} shows the resulting overall miss rate for all of the channels specified for \ac{EC-GSM-IoT}. \SI{90}{\percent} of the simulations succeed at \iac{SNR} of \SI{-11.5}{\decibel} in the low band \ac{ST} channel and at \iac{SNR} of \SI{-7.7}{\decibel} in the low band \ac{TU}1.2 channel. This exceeds the \ac{3GPP} goals by \SI{3}{\decibel} and \SI{2.2}{\decibel}, respectively. The \ac{CDF} for the synchronization time is shown in Fig.~\ref{fig:Synchronization time CDF}. It takes \SIrange{1.6}{1.8}{\second} to achieve \SI{90}{\percent} successful synchronizations, meeting Requirement~R1. The average synchronization time in \iac{ST} channel at \SI{-8.5}{\decibel} is \SI{1.3}{\second}, corresponding to a \SI{56}{\percent} reduction, compared to \cite{weber2017sawless}. The \ac{MFD} takes \SIrange{250}{425}{\milli\second} on average, the rest of the time is spent receiving the \ac{EC-SCH}. This is good from an energy standpoint, since the receiver can be turned off most of the time during the \ac{EC-SCH} reception, recalling the \ac{MF} structure in Fig.~\ref{fig:EC-GSM-IoT Multiframe}.

\begin{figure}[!ht]
\begin{center}
\includegraphics[width=\figurewidth]{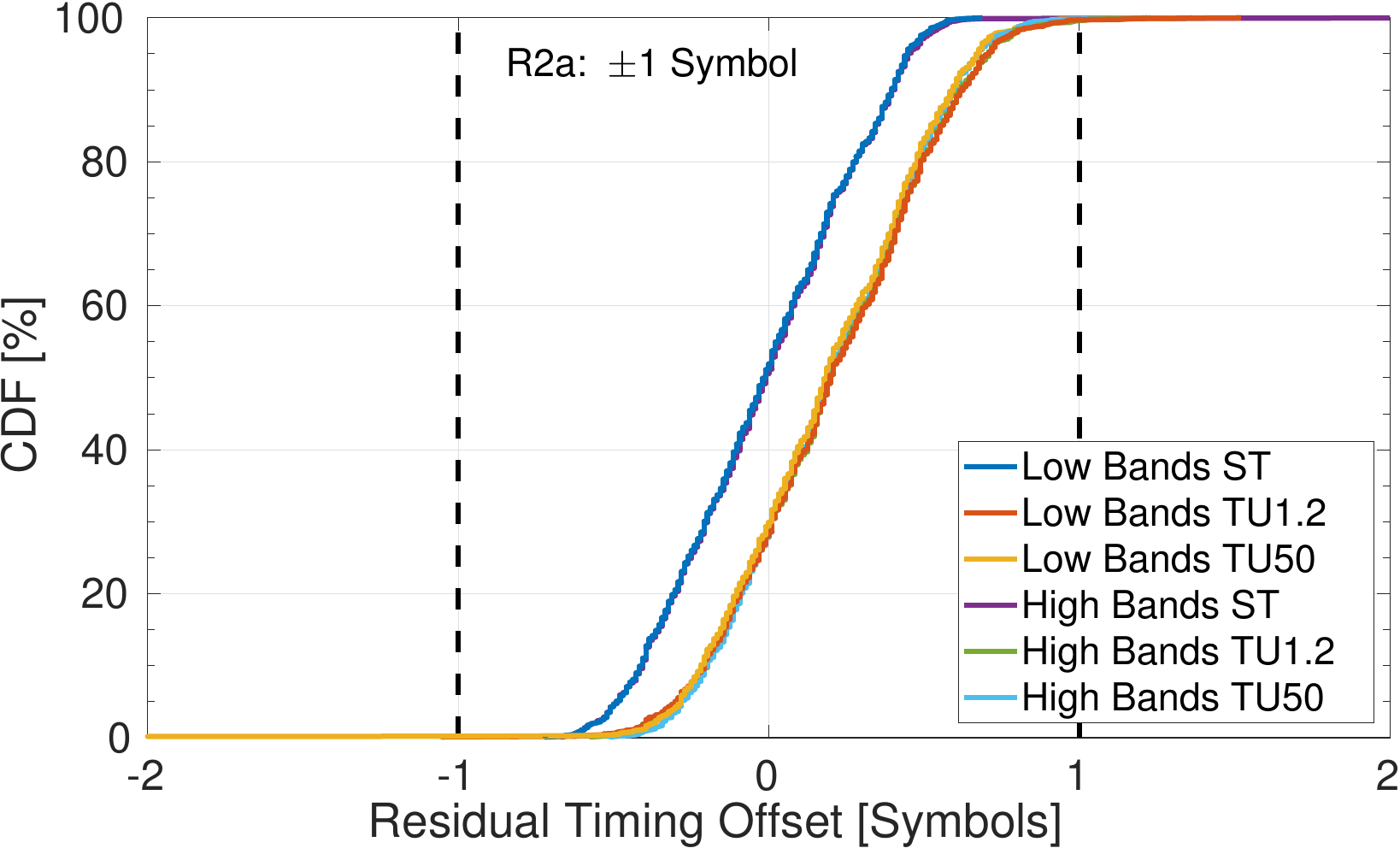}
\caption[Synchronization TO CDF]{Residual time offset \acs*{CDF} at the \acs*{ISL} for \acs*{EC-SCH} reference performance, assuming \iac*{NF} of \SI{5}{\decibel}.}
\label{fig:Synchronization TO CDF}
\end{center}
\end{figure}

\begin{figure}[!ht]
\begin{center}
\includegraphics[width=\figurewidth]{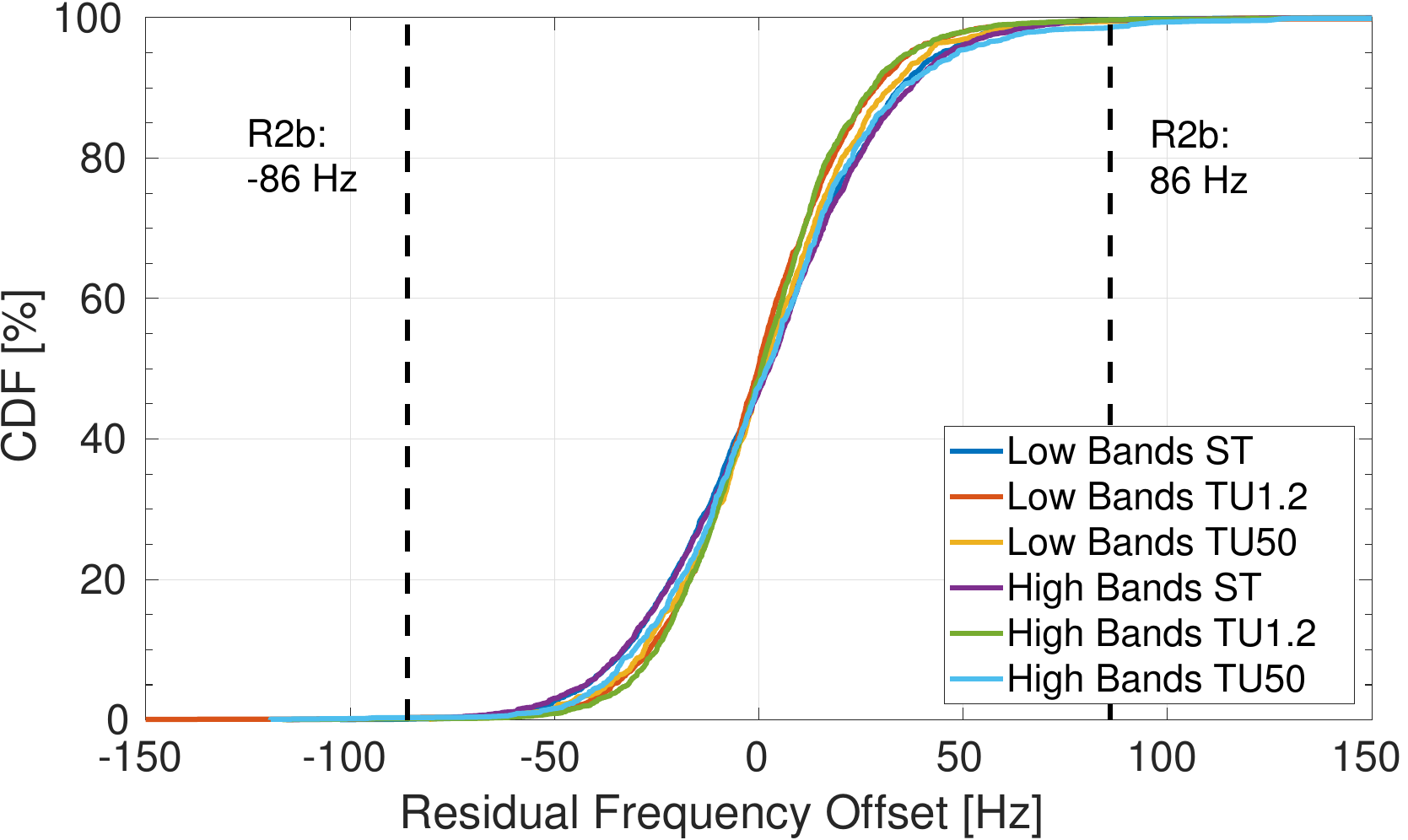}
\caption[Synchronization FO CDF]{Residual frequency offset \ac*{CDF} at the \ac*{ISL} for \ac*{EC-SCH} reference performance, assuming \iac*{NF} of \SI{5}{\decibel}. Unsuccessful synchronizations are counted towards to total, but not shown.}
\label{fig:Synchronization FO CDF}
\end{center}
\end{figure}

Fig.~\ref{fig:Synchronization TO CDF} shows the fine time offset at the end of the synchronization process. It is below \num{\pm 1} symbol in more than \SI{99.5}{\percent} of the attempts for all the channels at the \ac{ISL} for \ac{EC-SCH} reference performance, which meets Requirement~R2a. As shown in Fig.~\ref{fig:Synchronization FO CDF}, the residual frequency offset after the fine \ac{FOE} over 40 \acp{FB} is below \SI{0.1}{\ppm} in \SI{99}{\percent} of the test cases, which meets Requirement~R2b.

\section{Cell Search}
\label{sec:cell search}

A cell search may have to be performed, whenever the device cannot reach \iac{BS} at a known frequency. Then, it has to scan all $298+673$ potential low and high band \ac{BCCH} carriers. Legacy \ac{GSM} devices look for a suitable carrier by performing receive power measurements, which are not feasible below the noise floor. The duration of the cell search becomes problematic in \ac{EC-GSM-IoT}, because the device must perform a partial synchronization to all potential carriers \cite{3gpp:gp160091}. Like other \ac{cIoT} standards, \ac{EC-GSM-IoT} trades synchronization time for an improvement in coverage. As in \cite{3gpp:gp160091}, we propose the use of the \ac{MFD} to detect the presence of \iac{GSM} carrier in extended coverage. Their algorithm requires 10 \acp{MF} to achieve a \SI{99}{\percent} detection rate in \iac{TU}\num{1.2} channel. Our proposed algorithm can achieve the same performance after searching only 4 \acp{MF}. However, the limiting case is the high band \ac{TU}50 channel with a maximum frequency offset of \SI{\pm 50}{\kilo\hertz}. In this case, 7 \acp{MF} are required to achieve a \SI{99}{\percent} detection rate. No false detections were simulated in \num{10000} test cases. With this \SI{30}{\percent} reduction compared to \cite{3gpp:gp160091}, the cell search takes \num{27} minutes and a maximum of \num{113} full cell searches are possible with a \SI{5}{\watt\hour} battery \footnote{Assuming a total receiver power consumption of \SI{100}{\milli\watt} \cite{3gpp:45820}. Note that the noise bandwidth definition in \cite{3gpp:gp160091} differs, and results in \iac{SNR} offset of \SI{1.3}{\decibel}.}.


\section{Conclusion}

We have presented the first complete set of algorithms, which is able to successfully synchronize at \iac{SNR} as low as \SI{-11.5}{\decibel}. It fulfills the requirements for the synchronization time and the residual time and frequency offset with a low computational complexity. Our simulations show that it exceeds the \ac{3GPP} target by \SI{3}{\decibel} in the best-case, resulting in \iac{MCL} of up to \SI{170.5}{\decibel}. In the worst-case the target is exceeded by \SI{2.2}{\decibel}, corresponding to \iac{MCL} of \SI{166.7}{\decibel}. The network synchronization supports a large frequency offset oscillator, enabling a low-cost \ac{EC-GSM-IoT} implementation.


\section*{Acknowledgment}
We would like to thank Innosuisse for their financial support through project 18987.1PFNM-NM.

\newpage

%% file: acronyms.tex
\acresetall

\begin{acronym}[AAA]
\acro{LTE}{Long Term Evolution}\acroindefinite{LTE}{an}{a}
\acro{4G}{4th Generation}
\acro{5G}{5th Generation}
\acro{LPWAN}{Low-Power Wide Area Network}
\acro{M2M}{Machine to Machine}\acroindefinite{M2M}{an}{a}
\acro{IoT}{Internet of Things}\acroindefinite{IoT}{an}{an}
\acro{NB-IoT}{Narrow-Band IoT}\acroindefinite{NB-IoT}{an}{a}
\acro{eMTC}{enhanced MTC}\acroindefinite{eMTC}{an}{an}
\acro{RF}{Radio Frequency}\acroindefinite{RF}{an}{a}
\acro{SoC}{System on Chip}
\acro{RF-SoC}{}\acroindefinite{RF-SoC}{an}{a}
\acro{DBB}{Digital Baseband}
\acro{DEC}{DECoder}
\acro{DET}{DETector}
\acro{DFE}{Digital Front End}
\acro{FMC}{FPGA Mezzanine Card}\acroindefinite{FMC}{an}{an}
\acro{LPC}{Low Pin Count}\acroindefinite{LPC}{an}{a}
\acro{HPC}{High Pin Count}\acroindefinite{HPC}{an}{a}
\acro{RBDP}{Radio Front End - Baseband Digital Parallel}\acroindefinite{RBDP}{an}{a}
\acro{FPGA}{Field Programmable Gate Array}\acroindefinite{FPGA}{an}{a}
\acro{DSP}{Digital Signal Processor}
\acro{CPU}{}
\acro{VLSI}{}
\acro{MIMO}{}
\acro{ASIC}{Application Specific Integrated Circuit}\acroindefinite{ASIC}{an}{an}
\acro{SMA}{SubMiniature version A}\acroindefinite{SMA}{an}{a}
\acro{IC}{Integrated Circuit}\acroindefinite{IC}{an}{an}
\acro{GMSK}{Gaussian Minimum Shit Keying}
\acro{QAM}{Quadrature Amplitude Modulation}
\acro{FER}{Frame Erasure Rate}
\acro{BLER}{BLock Error Rate}
\acro{2G}{2nd Generation}
\acro{3G}{3rd Generation}
\acro{3GPP}{3rd Generation Partnership Project}
\acro{SDR}{Software Defined Radio}\acroindefinite{SDR}{an}{a}
\acro{HDL}{Hardware Description Language}\acroindefinite{HDL}{an}{a}
\acro{CMOS}{}
\acro{NF}{Noise Figure}\acroindefinite{NF}{an}{a}
\acro{VHF}{Very High Frequency}
\acro{UE}{User Equipment}
\acro{SAW}{Surface Acoustic Wave}
\acro{SPI}{Serial Peripheral Interface Bus}\acroindefinite{SPI}{an}{a}
\acro{MAC}{Medium Access Control}
\acro{TMU}{Time Management Unit}
\acro{MCL}{Maximum Coupling Loss}\acroindefinite{MCL}{an}{a}
\acro{LLR}{Log-Likelihood Ratio}\acroindefinite{LLR}{an}{a}
\acro{DL}{Down-Link}
\acro{ISI}{Inter Symbol Interference}\acroindefinite{ISI}{an}{an}
\acro{PCB}{Printed Circuit Board}
\acro{PA}{Power Amplifier}
\acro{CC}{Coverage Class}
\acro{SNR}{Signal to Noise Ratio}\acroindefinite{SNR}{an}{a}
\acro{BEP}{Bit Error Probability}
\acro{MCS1}{Modulation and Coding Scheme 1}\acroindefinite{MCS1}{an}{a}
\acro{CS1}{Coding Scheme 1}
\acro{EC-MCS1}{Extended-Coverage Modulation and Coding Scheme 1}\acroindefinite{EC-MCS1}{an}{an}
\acro{AWGN}{Additive White Gaussian Noise}\acroindefinite{AWGN}{an}{an}
\acro{IQ}{In-phase and Quadrature}\acroindefinite{IQ}{an}{an}
\acro{Rx}{Receiver}\acroindefinite{Rx}{an}{a}
\acro{Tx}{Transmitter}
\acro{PAN}{Personal Area Network}
\acro{WLAN}{Wireless Local Area Network}
\acro{RBER}{Residual Bit Error Rate}\acroindefinite{RBER}{an}{a}
\acro{HD}{Half Duplex}\acroindefinite{HD}{an}{a}
\acro{FDD}{Frequency Division Duplex}\acroindefinite{FDD}{an}{a}
\acro{VD}{Viterbi Decoder}
\acro{TDMA}{Time Division Multiple Access}
\acro{CORDIC}{COordinate Rotation DIgital Computer}
\acro{SRAM}{Static Random Access Memory}\acroindefinite{SRAM}{an}{a}
\acro{TTI}{Transmission Time Interval}
\acro{RMS}{Root Mean Square}\acroindefinite{RMS}{an}{a}
\acro{TS}{TimeSlot}\acroindefinite{TS}{a}{a}
\acro{CRC}{Cyclic Redundancy Check}\acroindefinite{CRC}{a}{a}
\acro{CDF}{Cumulative Distribution Function}\acroindefinite{CDF}{a}{a}
\acro{ISL}{Input Signal Level}\acroindefinite{ISL}{an}{an}
\acro{CRLB}{Cramér-Rao Lower Bound}\acroindefinite{CRLB}{a}{a}
\acro{DFT}{Discrete Fourier Transform}\acroindefinite{DFT}{a}{a}
\acro{FFT}{Fast Fourier Transform}\acroindefinite{FFT}{an}{a}
\acro{cIoT}{cellular \ac{IoT}}\acroindefinite{cIoT}{a}{a}
\acro{RSSI}{Received Signal Strength Indicator}\acroindefinite{RSSI}{an}{a}
\acro{BS}{Base Station}\acroindefinite{BS}{a}{a}
\acro{CTI}{Swiss Commission for Technology and Innovation}

\acro{GPRS}{}
\acro{GSM}{}
\acro{EGPRS2A}{}
\acro{EDGE}{}\acroindefinite{EDGE}{an}{an}
\acro{EC-GSM-IoT}{Extended-Coverage \ac{GSM} for the \ac{IoT}}\acroindefinite{EC-GSM-IoT}{an}{an}

\acro{FCCH}{Frequency Correction CHannel}\acroindefinite{FCCH}{an}{a}
\acro{SCH}{Synchronization CHannel}\acroindefinite{SCH}{an}{a}
\acro{EC-SCH}{Extended-Coverage \ac{SCH}}\acroindefinite{EC-SCH}{an}{an}
\acro{BCCH}{Broadcast Control CHannel}\acroindefinite{BCCH}{a}{a}
\acro{EC-BCCH}{Extended Coverage \ac{BCCH}}\acroindefinite{EC-BCCH}{a}{a}
\acro{PDTCH}{Packet Data Traffic CHannel}
\acro{EC-PDTCH}{Extended-Coverage PDTCH}\acroindefinite{EC-PDTCH}{an}{an}

\acro{MFD}{\ac{MF} boundary Detection}\acroindefinite{MFD}{an}{an}
\acro{FBD}{\ac{FB} Detection}\acroindefinite{FBD}{an}{an}
\acro{SB}{Synchronization Burst}\acroindefinite{SB}{an}{a}
\acro{ML}{Maximum Likelihood}\acroindefinite{ML}{an}{a}
\acro{SOVE}{Soft-Output Viterbi Equalizer}
\acro{MF}{Multi-Frame}\acroindefinite{MF}{an}{a}
\acro{FB}{Frequency Burst}\acroindefinite{FB}{an}{a}
\acro{FOE}{Frequency Offset Estimation}\acroindefinite{FOE}{an}{a}
\acro{ST}{STatic}\acroindefinite{ST}{an}{a}
\acro{TU}{Typical Urban}\acroindefinite{TU}{a}{a}


\acro{NR}{New Radio}\acroindefinite{NR}{an}{a}

\end{acronym}

\acused{IoT}
\acused{EDGE}
\acused{EC-GSM-IoT}
\acused{LTE}
\acused{4G}
\acused{5G}
\acused{NB-IoT}
\acused{eMTC}
\acused{PAN}
\acused{WLAN}
\acused{GSM}
\acused{GPRS}
\acused{EGPRS}
\acused{EGPRS2A}
\acused{RF-SoC}
\acused{SNR}
\acused{AWGN}
\acused{IQ}
\acused{BLER}
\acused{VLSI}
\acused{HD}
\acused{FDD}
\acused{QAM}
\acused{RF}
\acused{SAW}
\acused{SPI}
\acused{TDMA}
\acused{PCB}
\acused{PA}
\acused{EC-MCS1}
\acused{FPGA}
\acused{MCS1}
\acused{3GPP}
\acused{RBER}
\acused{Rx}
\acused{CORDIC}
\acused{CMOS}
\acused{SRAM}
\acused{CS1}
\acused{GMSK}
\acused{DFT}
\acused{FFT}
\acused{MIMO}